\documentclass[3p,times,procedia]{elsarticle}
\usepackage{nupha_ecrc}
\usepackage{hyperref}
\usepackage{lineno}

\usepackage{graphicx}
\usepackage[loose]{units}
\usepackage{upgreek}
\usepackage{color}
\usepackage[usenames,dvipsnames]{xcolor}
\usepackage{amsmath,amssymb,amsbsy} 

\volume{00}

\firstpage{1}

\journalname{Nuclear Physics A}

\runauth{}

\jid{nupha}

\jnltitlelogo{Nuclear Physics A}

\usepackage{amssymb}

\usepackage[figuresright]{rotating}

\begin{document}

\newcommand{\PbPb}{\textnormal{Pb--Pb}}
\newcommand{\AuAu}{\textnormal{Au--Au}}
\newcommand{\pp}{\ensuremath{\mbox{p}\mbox{p}}}
\newcommand{\pt}{\ensuremath{p_\mathrm{T}}}
\newcommand{\pT}{\pt}
\newcommand{\ptch}{\ensuremath{p_\mathrm{T,\,ch\;jet}}}
\newcommand{\deltaptch}{\ensuremath{\delta p_\mathrm{T,\,ch}}}
\newcommand {\snnbf}{\ensuremath{\mathbf{\sqrt{s_{_{\mathrm{NN}}}} }}}
\newcommand {\snn}{\ensuremath{\sqrt{s_{_{\mathrm{NN}}}} }}

\newcommand{\dirg}       {\ensuremath{{\rm \gamma, dir}}}
\newcommand{\decg}       {\ensuremath{{\rm \gamma, dec}}}
\newcommand{\incg}       {\ensuremath{{\rm \gamma, inc}}}

\newcommand{\Rg}         {\ensuremath{R_{\rm \gamma}}}
\newcommand{\vdirg}      {\ensuremath{v_{2}^{\dirg}}}
\newcommand{\vdecg}      {\ensuremath{v_{2}^{\decg}}}
\newcommand{\vincg}      {\ensuremath{v_{2}^{\incg}}}
\newcommand{\vbckg}      {\ensuremath{v_{2}^{\rm \gamma, bck}}}

\begin{frontmatter}



\dochead{XXVIIth International Conference on Ultrarelativistic Nucleus-Nucleus Collisions\\ (Quark Matter 2018)}

\title{Direct photon elliptic flow\\ in Pb--Pb collisions at $\snn=2.76$ TeV}


\author{Mike Sas, for the ALICE Collaboration}

\address{University of Utrecht \& Nikhef, Netherlands}

\begin{abstract}
The elliptic flow of inclusive and direct photons was measured by ALICE for central and semi-central Pb--Pb collisions at $\snn=2.76$ TeV.
The photons were reconstructed using the electromagnetic calorimeter PHOS and the central tracking system. The inclusive photon flow reconstructed with both methods are combined and used to extract the direct photon flow, using a decay photon simulation and the direct photon excess $\Rg$, in the transverse momentum range $0.9 < \pT < \unit[6.2]{GeV}/c$.
We find agreement to earlier results obtained at RHIC, and the theoretical predictions generally under-predict the results.
\end{abstract}


\end{frontmatter}


\section{Introduction}
\label{Introduction}

Ultrarelativistic nucleus--nucleus collisions give access to experimentally study the Quark-Gluon Plasma (QGP) \cite{Borsanyi:2013lat,Bazavov:2012lat}, which is the main goal of the ALICE experiment. This droplet of hot QCD matter expands, cools down, and transforms into ordinary matter. The measured elliptic flow of final state particles has been interpreted as being due to collective expansion, which transforms the initial spatial anisotropy into a momentum anisotropy of the final state particles.
Here, the emphasis is on the elliptic flow of direct photons. Direct photons are the photons not coming from hadronic decays, and are produced during all stages of the collision. Direct photons are a unique tool to probe the QGP since they leave the system without interacting with the formed medium, because their mean free path is much larger than the size of the system. At low transverse momentum, the direct photons are mainly thermally produced by the hot matter, and their production rate can be used to estimate the temperature of the medium. Additionally, their elliptic flow provides information on the development of flow during the whole evolution of the system.
Collective flow is quantified by the azimuthal distribution of particles, which is expanded as $1+2 \sum v_{n} \cos[n (\varphi-\Psi_{RP})]$ \cite{Voloshin:1994ph}, where $\varphi$ is the azimuthal angle of the measured particle, and $\Psi_{RP}$ is the reaction plane orientation.
The theoretical calculations are predicting a smaller elliptic flow of direct photons compared to that of hadrons. This is because the direct photons probe the momentum anisotropy at the time of their emission, which occurs when the system is still expanding and developing the final momentum anisotropy.
The measurement of inclusive and direct photon elliptic flow in Pb--Pb collisions is presented, and compared to results at RHIC as well as to theoretical calculations. More details on the analysis and results can be found in \cite{Alicedirgv2:2018}.

\section{Analysis method}
\label{Method}

The presented results use the \PbPb\ data recorded by the ALICE experiment in 2010. For the measurement of the inclusive photon flow, two methods are used to reconstruct the photons: the Photon Conversion Method (PCM) and the calorimeter PHOS \cite{Dellacasa:1999kd}. In the PCM, the Inner Tracking System (ITS) and the Time Projection Chamber (TPC) are used to reconstruct the $e^+e^-$ pairs, which result from a photon converting in the detector material. For PHOS, the photons are reconstructed by measuring their energy deposit in the calorimeter. The scintillator array detectors V0A and V0C are used for the minimum bias trigger and the event plane orientation calculation, which subtend the pseudorapidity ranges $2.8 < \eta < 5.1$ and $-3.7 < \eta < -1.7$, respectively. Furthermore, the events are divided into two centrality classes; central (0--20\%) and semi-central (20--40\%), according to the summed amplitudes of the V0A and V0C detectors.

The direct photon elliptic flow is calculated using
\begin{equation}
  \vdirg=\frac{\vincg \Rg - \vdecg }{\Rg-1},
  \label{vdir}
\end{equation}
where $\vincg$ is the inclusive photon flow, $\vdecg$ the decay photon flow, and $\Rg = N_{\incg}/N_{\decg}$ is the ratio quantifying the excess of direct photons. The ratio $\Rg$ was measured and can be found in \cite{Adam:2015lda}. The decay photon flow is estimated by a simulation, using the existing measurements of hadronic flow. Furthermore, the inclusive photon elliptic flow is measured with the Scalar Product method \cite{Adler:2002pu}, using a pseudorapidity gap of $|\Delta\eta| > 0.9$ between the photon and the reference flow particles. The results from the PCM and PHOS measurement are combined and treated as independent.

\begin{figure}[t]
\unitlength\textwidth
\centering
\includegraphics[width=0.49\linewidth]{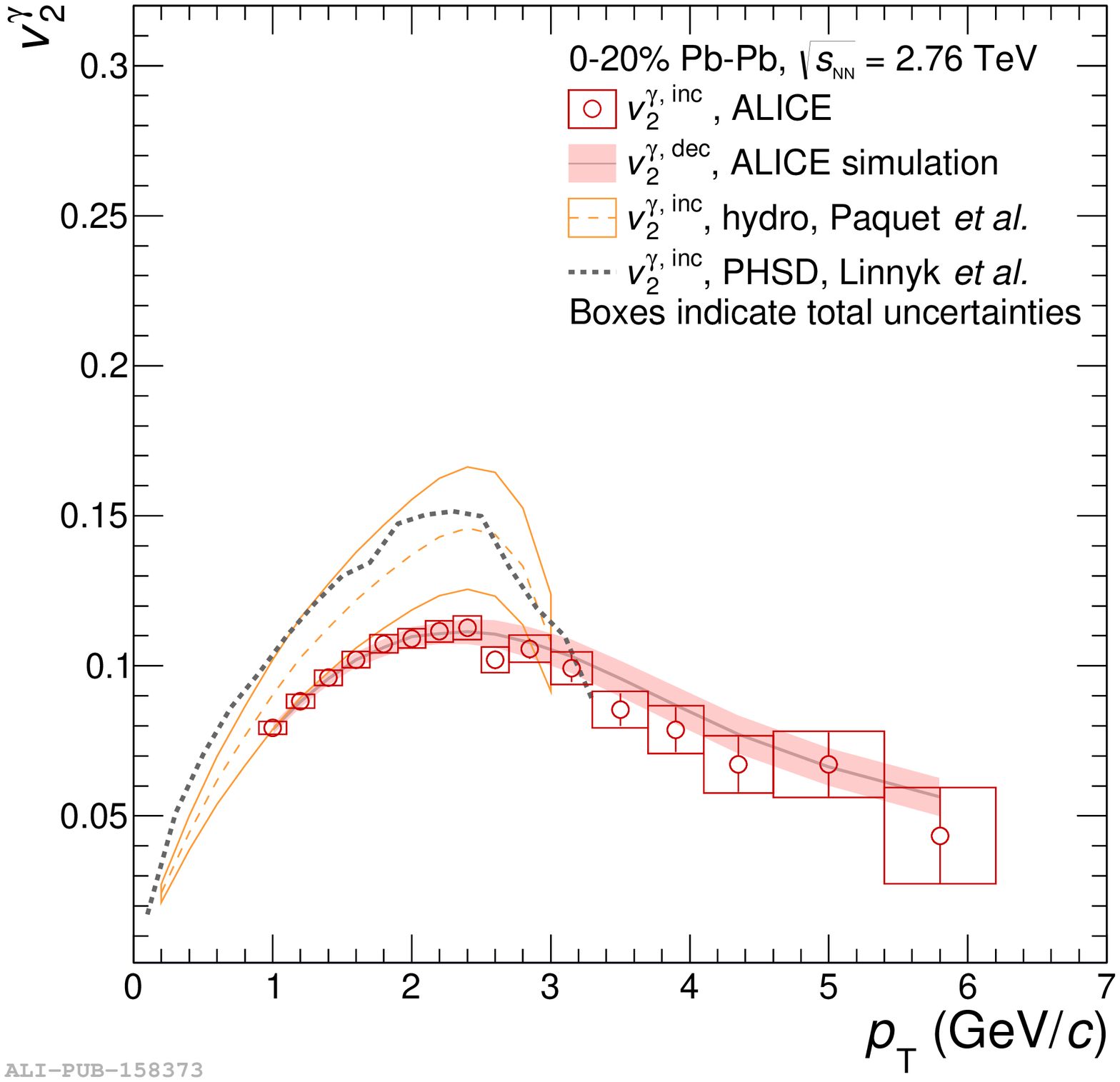}
\hfill
\includegraphics[width=0.49\linewidth]{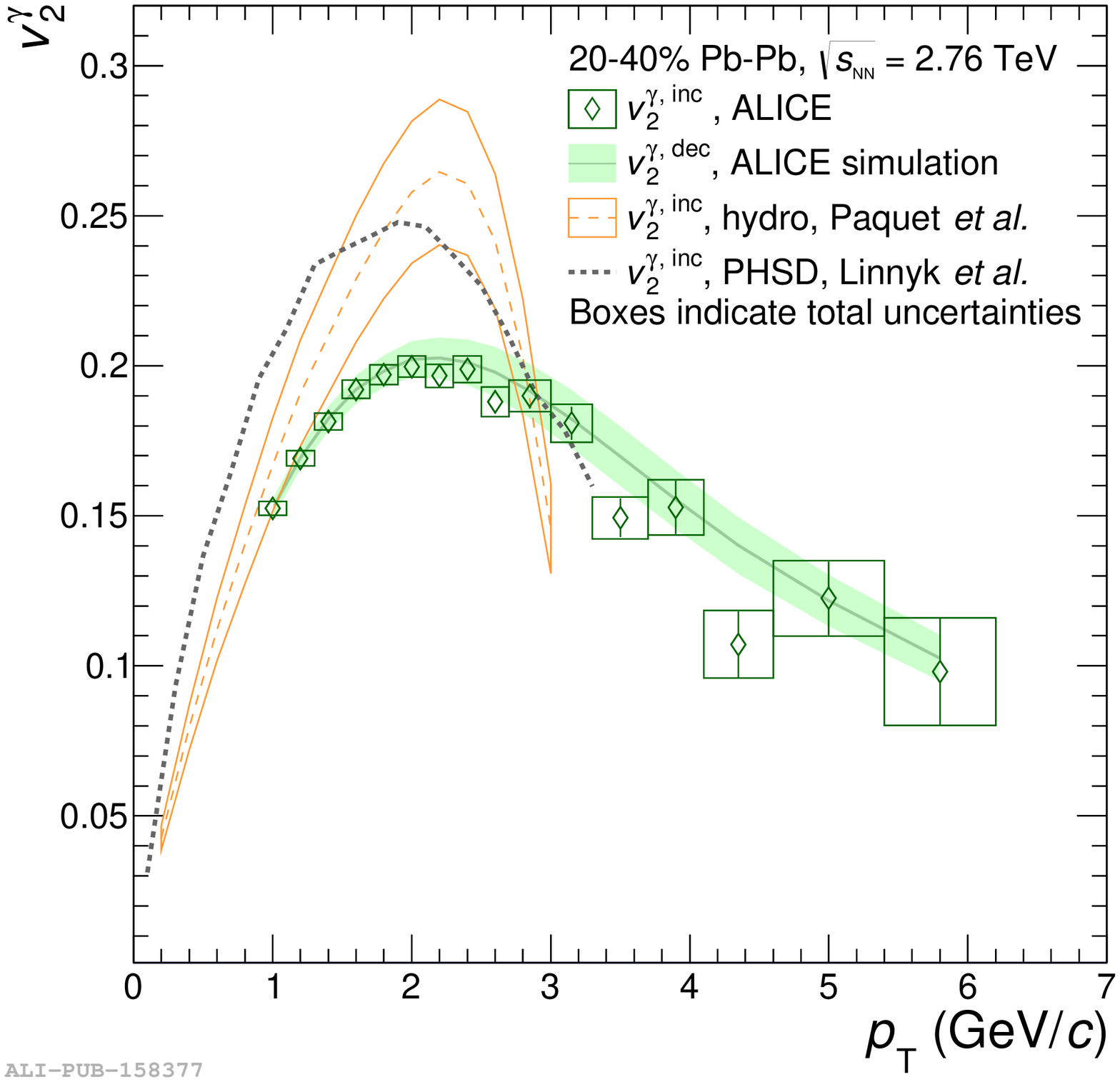}
\caption{(Color online) The elliptic flow of inclusive photons and decay photons in Pb--Pb collisions at $\snn=2.76$ TeV, compared to hydrodynamic \cite{Gale:2014dfa} and transport PHSD \cite{Linnyk:2013wma} model predictions, for the 0--20\% (left) and 20--40\% (right) centrality classes. The boxes indicate the total uncertainty, and the vertical bars indicate the statistical uncertainty.
\label{fig:Inclv2Theory}}
\end{figure}

\section{Results}
\label{Results}

The inclusive and decay photon elliptic flow, as measured in two centrality classes, is shown in Fig.~\ref{fig:Inclv2Theory}.
Over the full transverse momentum range of $0.9<\pT<6.2$~GeV/$c$, the flow of inclusive and decay photons are similar, which results from the fact that the inclusive photon flow is dominated by the decay photons. Interestingly, the theoretical prediction described in \cite{Gale:2014dfa} and \cite{Linnyk:2013wma} overshoots the data by about 40\% in the transverse momentum range $1<\pT<3$~GeV/$c$, showing that the identified hadron flow is not fully described in the given models.

\begin{figure}[ht]
\unitlength\textwidth
\centering
\includegraphics[width=0.49\linewidth]{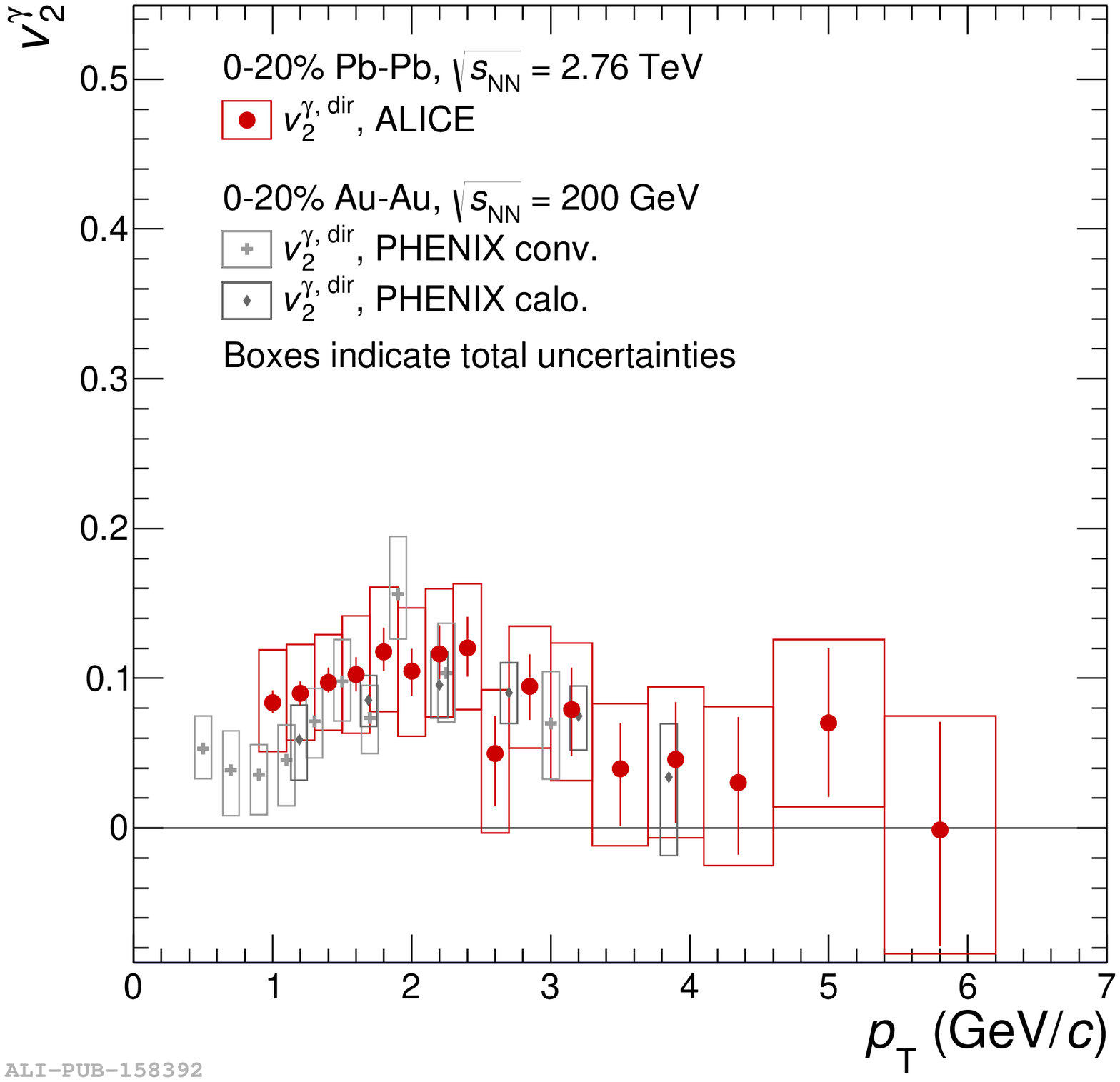}
\hfill
\includegraphics[width=0.49\linewidth]{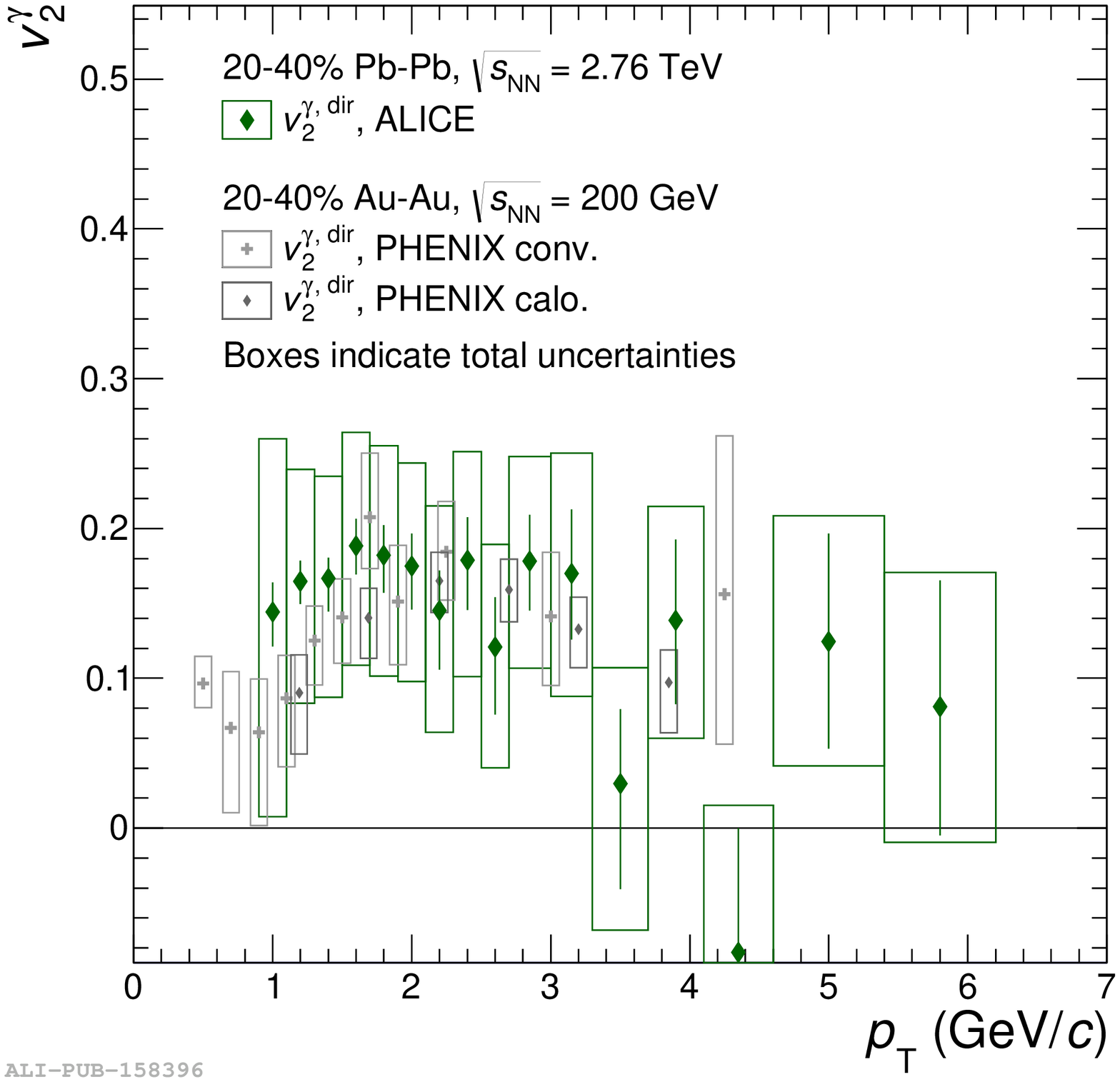}
\caption{(Color online) The elliptic flow of direct photons in Pb--Pb collisions at $\snn=2.76$ TeV, compared with PHENIX results \cite{Adare:2015lcd} in Au--Au collisions at $\snn=200$ GeV, for the 0--20\% (left) and 20--40\% (right) centrality classes. The boxes indicate the total uncertainty, and the vertical bars indicate the statistical uncertainty.
\label{fig:v2DirPHENIX}}
\end{figure}

\begin{figure}[ht]
\unitlength\textwidth
\centering
\includegraphics[width=0.49\linewidth]{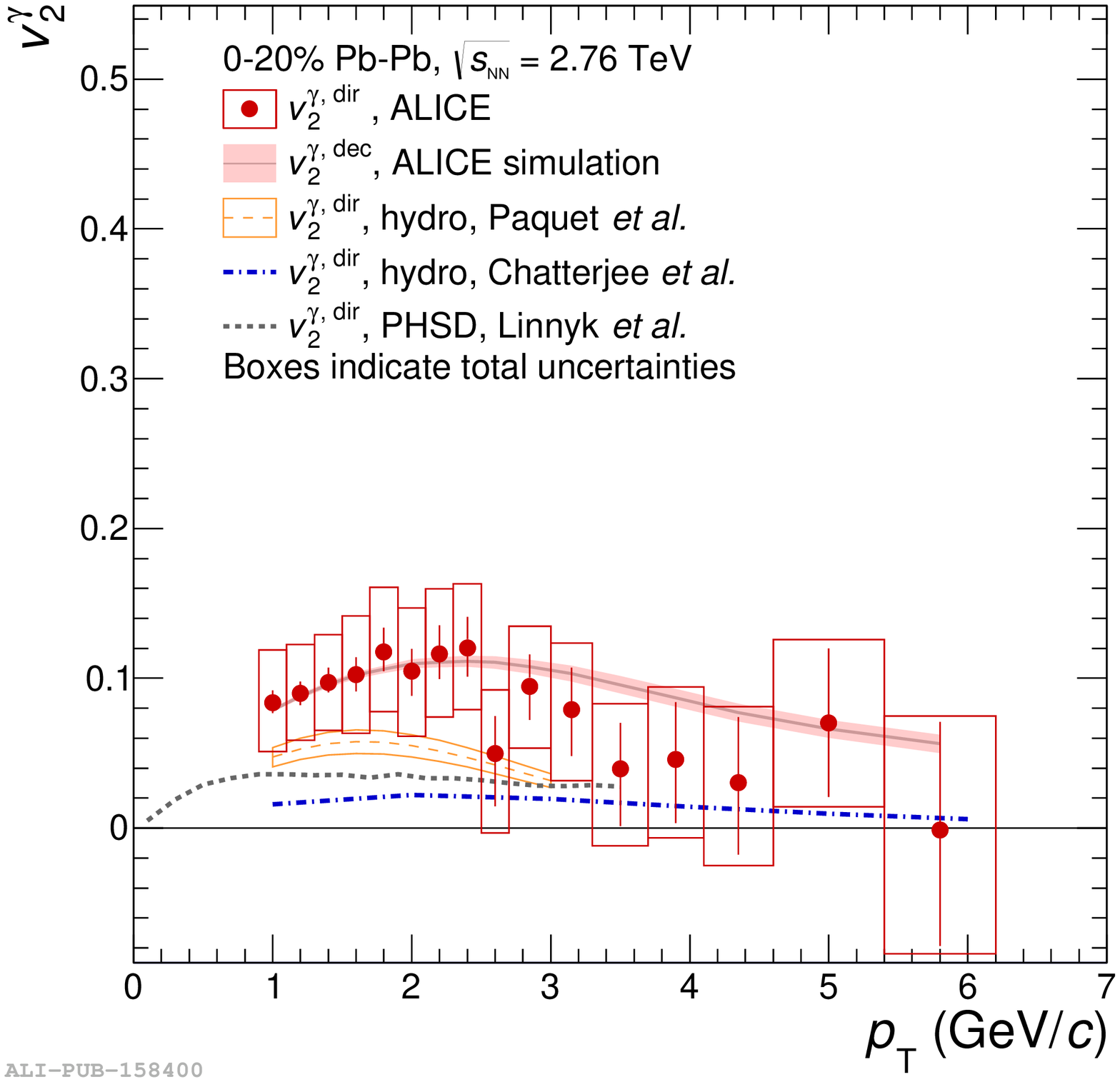}
\hfill
\includegraphics[width=0.49\linewidth]{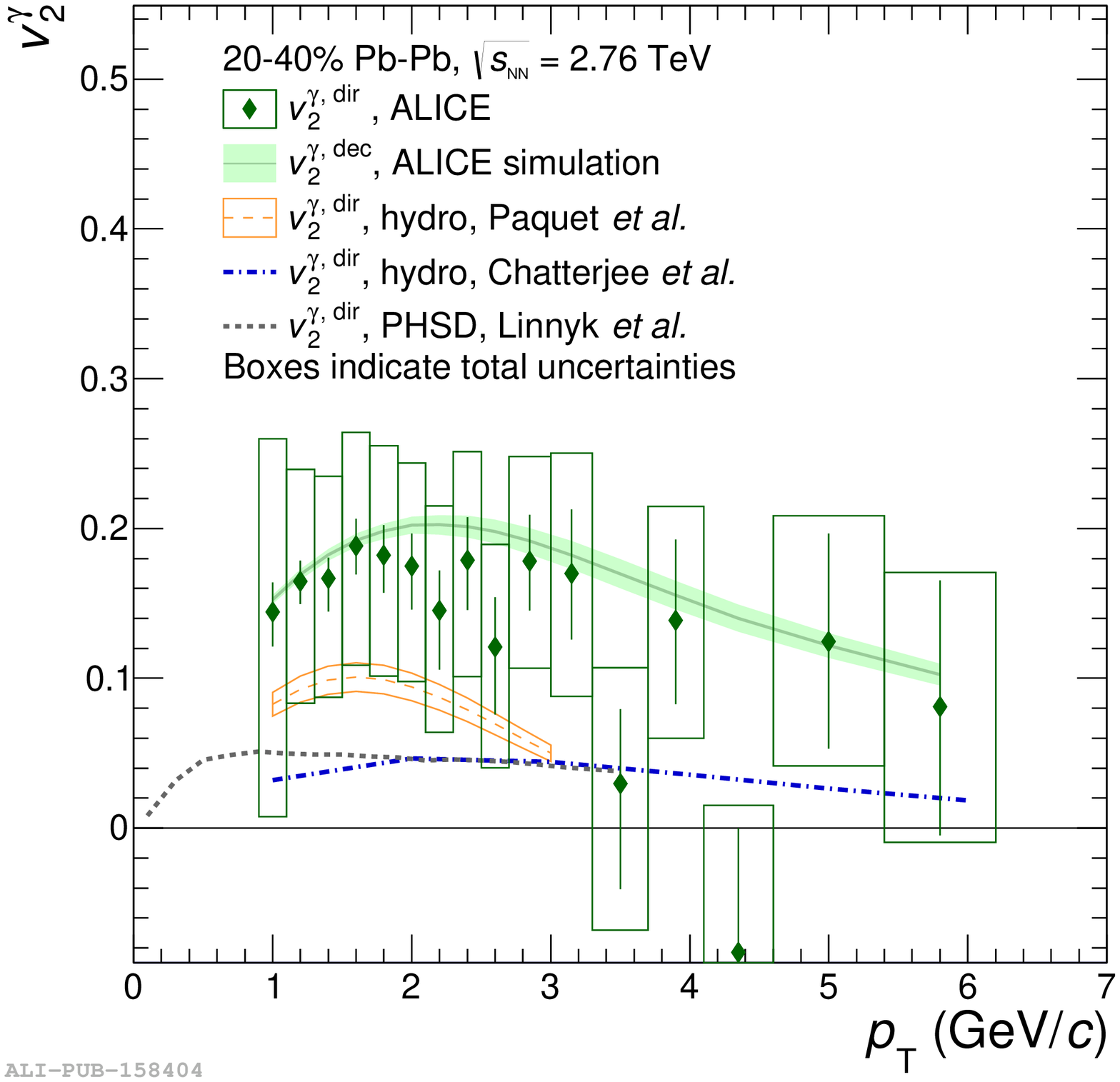}
\caption{(Color online) The elliptic flow of direct photons in Pb--Pb collisions at $\snn=2.76$ TeV, compared to hydrodynamic \cite{Gale:2014dfa} and transport PHSD \cite{Linnyk:2013wma} model predictions in the 0--20\% (left) and 20--40\% (right) centrality classes. The boxes indicate the total uncertainty, and the vertical bars indicate the statistical uncertainty.
\label{fig:v2DirTHEORY} }
\end{figure}

As given by Eq. \ref{vdir}, the direct photon $v_2$ is calculated by subtracting the decay photon flow from the inclusive photon flow, using the direct photon excess $\Rg$.
Since the significance of $\Rg>1$ is limited, it was chosen to use a Bayesian approach to extract the direct photon elliptic flow. The true value of $\Rg$ is restricted to be greater than unity, corresponding to a constant prior for $R_\mathrm{\gamma,true} \ge 1$. The posterior distributions are sampled for each $\pT$ bin, from which $\vdirg$ is calculated.
The resulting direct photon elliptic flow are shown in Fig.~\ref{fig:v2DirPHENIX} and Fig.~\ref{fig:v2DirTHEORY}, for both centrality classes 0--20\% and 20--40\%, respectively. The total uncertainty is represented by the boxes, and the error bars represent the statistical uncertainty. In Fig.~\ref{fig:v2DirPHENIX} the results are compared to earlier measurements performed at RHIC energies by the PHENIX collaboration \cite{Adare:2015lcd}. Both results are found to be compatible within uncertainties. In Fig.~\ref{fig:v2DirTHEORY} data are compared to theoretical calculations using state-of-the-art hydrodynamic model \cite{Gale:2014dfa,Chatterjee:2017akg} and the PHSD transport model \cite{Linnyk:2015tha}. The calculations in general under-predict the data, but the difference is not significant due to the large experimental uncertainties.

\section{Summary}
\label{Summary}

As a summary, The measurement of inclusive and direct photon elliptic flow for central and semi-central Pb--Pb collisions at $\snn=2.76$ TeV are presented. The results of the inclusive photon elliptic flow are the combination of a photon conversion and calorimetric measurement, where both measurements cover the transverse momentum range of $0.9<\pT<6.2$~GeV/$c$. To extract the direct photon elliptic flow, the decay photon flow is subtracted from the inclusive photon elliptic flow, using the previously measured direct photon excess $\Rg$.
$\vincg$ shows to be similar to $\vdecg$, which is expected since the inclusive photon sample is dominated by the photons coming from hadronic decays. However, theoretical prediction are over-predicting the data. In addition, $\vdirg$ also appears to be close to $\vdecg$, confirming earlier results at RHIC energies. The theoretical predictions for the direct photon elliptic flow are, unlike the the prediction for inclusive photons, under-predicting the data. However, the uncertainties on the measurements are still too large in order to draw strong conclusions. Future measurements, utilizing a much larger statistics dataset, will significantly improve the precision of the measurement.





\bibliographystyle{elsarticle-num}
\bibliography{biblio.bib}







\end{document}